\documentclass{article}
\usepackage{graphicx}

\newcommand{\bA}{\mbox {\bf A}}
\newcommand{\bc}{\mbox {\bf c}}
\newcommand{\be}{\mbox {\bf e}}
\newcommand{\bE}{\mbox {\bf E}}
\newcommand{\mE}{\mbox {E}}
\newcommand{\bH}{\mbox {\bf H}}
\newcommand{\bI}{\mbox {\bf I}}
\newcommand{\bp}{\mbox {\bf p}}
\newcommand{\bP}{\mbox {\bf P}}
\newcommand{\bQ}{\mbox {\bf Q}}
\newcommand{\bR}{\mbox {\bf R}}
\newcommand{\bs}{\mbox {\bf s}}
\newcommand{\bt}{\mbox {\bf t}}
\newcommand{\bu}{\mbox {\bf u}}
\newcommand{\bw}{\mbox {\bf w}}
\newcommand{\bx}{\mbox {\bf x}}
\newcommand{\by}{\mbox {\bf y}}
\newcommand{\bX}{\mbox {\bf X}}
\newcommand{\bY}{\mbox {\bf Y}}
\newcommand{\bZ}{\mbox {\bf Z}}
\newcommand{\bzer}{\mbox {\bf 0}}
\newcommand{\bone}{\mbox {\bf 1}}
\newcommand{\diag}{\mbox {diag}}
\newcommand{\var}{\mbox {var}}
\newcommand{\tra}{^{\top}}
\newcommand{\bbeta}{{\mbox {\boldmath$\beta$}}}
\newcommand{\bdelta}{{\mbox {\boldmath$\delta$}}}
\newcommand{\bgamma}{{\mbox {\boldmath$\gamma$}}}
\newcommand{\bGamma}{{\mbox {\boldmath$\Gamma$}}}
\newcommand{\bepsi}{{\mbox {\boldmath$\varepsilon$}}}
\newcommand{\blambda}{{\mbox {\boldmath$\lambda$}}}
\newcommand{\bpi}{{\mbox {\boldmath$\pi$}}}
\newcommand{\bchi}{{\mbox {\boldmath$\chi$}}}
\newcommand{\htheta}{{\hat {\theta}}}
\newcommand{\bnu}{{\mbox {\boldmath$\nu$}}}
\newcommand{\refb}[1]{(\ref{#1})}

\def\deno{1 + p_{K+1\,} \bx_{K+1\,}\tra \bH^{-1} \bx_{K+1}}
\def\bRQ{\left ( \lambda_{2\,} \bI + \bQ \right )^{-1}}

\begin{document}

\baselineskip 21pt 

\title{Stability of relaxed calibration}

\author{Nicholas T.\ Longford\footnote{SNTL Statistics Research and Consulting,
    3 Badgers Walk, Whyteleafe CR3 0AS, Surrey, UK.  
Email: {\tt sntlnick@sntl.co.uk}} \\[1mm]
  SNTL Statistics Research and Consulting, London, UK
  \\[1mm]
and Warsaw School of Economics (SGH), Warsaw, Poland}

\date{\empty} 
\maketitle

\thispagestyle{empty}
\setcounter{page}{0}
\vspace*{6mm}

\begin{abstract}

\noindent
Estimation of the population total of a variable can be improved
by calibration on a set of auxiliary variables.
It is difficult to establish that such a set of variables is sufficient,
that estimation could not be improved by calibration
on any further variables.
We address this issue by 
finding an upper bound for the change of the calibration estimate
of the population total of a variable when the auxiliary information
is supplemented by another variable 
for which the population total is known.
This upper bound can be interpreted as a measure of sensitivity
of the estimate to unavailable auxiliary information and considered
as a factor in deciding whether to seek further data sources
that would be included in calibration.
\end{abstract}


\clearpage

\section{Introduction} \label{S1}

Calibration of sampling weights is applied in many population surveys
that collect or have access to extensive auxiliary information.
Calibration has been the subject of intense interest ever since 
the seminal paper of Deville and S\"{a}rndal (1992);
see Haziza, and Beaumont (2017), Section 4, and Devaud and Till\'e (2019).
The original idea can be traced to Deming and Stephan (1940).

In this 
article
we pose the question
\begin{quote}
{\em How would the calibrated weights and the estimates based on them
change if additional auxiliary information became available?}
\end{quote}
and address it by constructing a variable for which the estimate
of the population total
would be
altered most within a plausible set
of auxiliary variables which we specify below. 
We believe that the related concern is well founded but has
not been attended to as much as its practical relevance deserves.
One reason for this inattention may have been the complexity
of the procedures for calibration, together with the dismissal
of any possible solution on the grounds that an additional
(henceforth {\em new}) auxiliary variable with known population total
could alter the calibrated weights and estimates based on them
to an arbitrary extent.

We propose a solution to this problem, qualified
by the largest plausible deviation of the estimate of the population total
of this new auxiliary variable from its true value
and the priority to reduce this deviation by re-calibration.
These two quantities,
the largest plausible deviation that would be subjected to re-calibration  
and the importance of the new variable,
are input parameters in the proposed algorithm.
The algorithm is an add-on to the method of relaxed calibration
(Longford, 2024), in which a priority parameter is specified
for each auxiliary variable.  The algorithm has several features
in common and can be motivated as an adaptation 
of the sensitivity analysis for hidden bias used in causal inference 
(Rosenbaum, 1995, Chapter 4; Rosenbaum and Silber, 2009).

The next section reviews the method of relaxed calibration 
and states the problem we address formally.
The solution is presented in Section \ref{Ssol}.
  Illustrative examples and the results of simulations
  are presented in Section \ref{Ssim}, with some details deferred
  to Supplementary Materials.
The concluding section presents a discussion and outlines
some unresolved issues and ideas for further research.

\section{Relaxed calibration} \label{Rev}

We assume that the survey data for a set of auxiliary variables are formed
in an $n \times (K+1)$ matrix $\bX$.  The columns of $\bX$ (variables)
are indexed 0, 1, \ldots, $K$ and the first column is the vector of unities,
denoted by $\bone_{n\,}$.  The subscript $n$ is dropped when its value
is immaterial or obvious from the context.
We use $\bzer$ for the vector of zeros and $\bI$ for the identity
matrix similarly.
The variables $\bx_{1\,}, \ldots, \bx_K$ in $\bX$ are standardized
  to have unweighted sample mean zero and unit variance;
  that is, $\bx_k\tra\bone = 0$ and $\bx_k\tra \bx_k = n$ for
  $k = 1, \ldots, K$.
Using $n$ or $n-1$ as the denominator for the variance is a matter
for convention; we use $n$ throughout.
The population totals of these variables,
  referred to as the targets,
are denoted by $\bt = (t_{0\,}, t_1, \ldots, t_k)\tra$;
they are assumed to be known.
  We make no assumptions about the population variances
  of these variables; usually they are not known.
The population size is $t_0 = N$.
The design weights are denoted by $\bw$.

Standardising a variable $\bx^{\circ}$, with its target $t^{\circ}$,
is invariant with respect to linear transformations.
That is, by standardising $a \bx^{\circ} - b$ we obtain
$\bx = s_{\circ}^{-1}(\bx^{\circ} - \bar{x}^{\circ})$, where
$\bar{x}^{\circ}$ is the sample mean
and $s_{\circ}$ the sample standard deviation
of $\bx^{\circ}$.  The corresponding transformation of $t^{\circ}$
yields $t = \{at^{\circ} - bN - N(a\bar{x}^{\circ} - b)\}/(as_{\circ})
=(t^{\circ} - N\bar{x}^{\circ}) / s_{\circ\,}$.
In particular, standardising $\bx^{\circ}$ and $\bx^{\circ} - Nt^{\circ}$
yield the same values of $\bx$ and $t$.

The purpose of calibration is to adjust the weights $\bw$
to {\em calibration weights} $\bu$ so that the discrepancies
$\delta_k = \bx_{k\,}\tra\bu - t_k$ would become as small as possible;
each $\delta_k$ can be viewed as the error in estimating $t_{k\,}$.
Established approaches aim to achieve $\delta_k = 0$ or to satisfy
constraints $|_{\,}\delta_{k\,}| \le d_k$ for specified nonnegative constants
$d_{k\,}$, $k=0, \ldots, K$.
Usually $d_0$ is set to zero.

In relaxed calibration, the variables in $\bX$ are associated with priorities
$\bp = (p_{0\,}, \ldots, p_K)$, set by the analyst, which quantify the importance
or urgency to reduce the absolute discrepancies $|_{\,}\delta_{k\,}|$.
A further priority $R \in [0, 1]$ is defined to reflect the desire for small
alteration of $\bw$ in relation to small dispersion of the calibrated weights $\bu$.
All the constraints associated with the discrepancies are replaced by penalties
in the objective function
\begin{eqnarray} \label{H1s}
F(\bu; \bw) &=& \sum_{k=0}^K p_{k\,}\delta_k^2 \,+\,
R_{\,} (\bu - \bw ) \tra (\bu - \bw ) + (1 - R)_{\,} \bu \tra \bu
\nonumber \\[1mm] &=&
\bu\tra\bH\bu - 2 \bu\tra\bs + D \,,
\end{eqnarray}
where
\begin{eqnarray*}
  \bH &=& \bI_n + \bX \bP \bX\tra
\\[1mm]  
\bs &=& R\bw + (1 - R)_{\,} \frac{t_0}{n} \bone_n + \bX \bP \bt
\end{eqnarray*}
and $D = \bt\tra\bP\bt + R\bw\tra\bw + (1-R)_{\,} t_0^2/n$.
The solution, where $F$ attains its minimum, is $\bu^{\ast} = \bH^{-1}\bs$.

Relaxed calibration replaces constraints that are {\em individual}
to each variable with a {\em collective} goal characterised
by the objective function $F$.
Its advantages include closed-form expression for $\bu$,
a simple way of exploring the properties of $\bu$ and
$\bdelta = (\delta_{0\,}, \ldots, \delta_K)$ as functions of $\bp$
and a natural interpretation of the tuning parameters $\bp$ and $R$.
Being the solution of a quadratic equation, the calibrated weights are unique.
The calibration estimator $\htheta = \bu\tra\by$ can be expressed
as an adjusted generalized ridge regression estimator,
$$
\htheta \,=\, \left ( \bt + \bP^{-1} \bnu_R \right ) \tra 
\left ( \bP^{-1} + \bX_{\,} \tra \bX \right )^{-1} \bX\tra\by
   + \by\tra\bepsi_{\rm R\,}, 
$$
where $\bnu_{\rm R}$ and $\bepsi_{\rm R}$ are defined by the regression-like
orthogonal decomposition
$R\bw + n^{-1}(1-R)_{\,}t_{0\,} \bone_n \,=\, \bX\bnu_{\rm R} + \bepsi_{\rm R\,}$.

Kwon, Kim and Qiu (2025) introduce a much more general framework
for calibration.  The quadratic kernel we use corresponds to the squared
loss in their Table 1, which lists a number of alternatives related
to generalized entropy.   The authors discuss in which settings these
alternatives are particularly well suited.  In contrast to our approach,
they regard removal of bias as an imperative.  Therefore they exclude
the design weights from the objective function
and involve them in constraints.
We aim at a balance of the reduction of bias and sampling variance.

\section{New auxiliary variable} \label{Ssol}

Would it be worthwhile to supplement the auxiliary data by another variable,
with its population total known?
When obtaining such information entails little expense and effort,
the conclusions of Longford (2024) indicate that such a variable
should be included in the auxiliary data and calibrated on  ---
more is better, unless the variable is assigned excessive priority.
However, when some expense and effort are involved, this outlay
should be weighed against the potential for more efficient estimation.
The largest plausible change of the estimate,
which delimits the range of plausible scenarios,
can also be interpreted as an uncertainty about the estimand,
the population total $\theta$ of the focal variable $\by$, 
additional to the sampling variation.

Chambers, Skinner and Wang (1999) suggest that the details of calibration
should be set with the aim to minimize the mean squared error (MSE),
which translates in our setting to establishing the range of
plausible MSEs of the re-calibrated estimator of $\theta$.
This is an unmanageable task as we would have to consider
the joint probabilities of inclusion, $\pi_{ij\,}$,
in addition to the marginal probabilities $\pi_i = 1/u_{i\,}$
for all units $i$ in the sample.
Instead, regarding the dispersion of the weights as a proxy
for sampling variance (Potthoff, Manton and Woodbury, 1992),
we assign a priority, $1-R$, to the sum of squares of the weights,
$\bu\tra\bu$.  Reducing the discrepancies $\delta_k$
is related to bias reduction.

The new
(unobserved) auxiliary variable is denoted
by $\bx_{K+1}$ and its population total by $t_{K+1\,}$.
We use the subscript $\dagger$ for objects that include
or are based on this variable, in addition to the original
(observed)
variables in $\bX$.
Thus, $\bX_{\dagger} = (\bX, ~\bx_{K+1})$,
$\bH_{\dagger} = \bI + \bX_{\dagger\,} \bP_{\dagger\,} \bX\tra_{\dagger}$,
where $\bP_{\dagger} = \diag(\bp_{\dagger})$ with $\bp_{\dagger} = (p_{0\,},
p_{1\,}, \ldots, p_{K\,}, p_{K+1})$, and so on.
We refer to calibration based on $\bX_{\dagger}$ as {\em re-calibration}.
We derive an expression for the difference $\Delta\bu = \bu_{\dagger} - \bu$
between the calibrated weights that use $\bX_{\dagger}$ and $\bX$, 
which is amenable to a discussion of how $\bx_{K+1}$ influences this difference.
We highlight the simplicity of such an evaluation with relaxed calibration
owing to the closed form of $\bu$.

We assume that $\bx_{K+1}$ is standardised to have zero sample mean
and unit standard deviation, as are all the other variables in $\bX$,
and its population total $t_{K+1}$ is transformed accordingly.
Therefore $t_{K+1}$ is related to the deviation,
or the estimation error, that re-calibration aims to diminish.
The re-calibrated weights are
$\bu_{\dagger} = \bH_{\dagger}^{-1\,} \bs_{\dagger\,}$, where 
\begin{eqnarray*} 
  \bH_{\dagger} &=& \bH \,+\, p_{K+1\,} \bx_{K+1\,} \bx_{K+1}\tra
\nonumber \\[1mm]   
\bs_{\dagger} &=& \bs \,+\, p_{K+1\,} t_{K+1\,} \bx_{K+1} \,,
\end{eqnarray*}
and $\bH$ and $\bs$ are given by equation \refb{H1s}.  Hence
{\arraycolsep 2.88pt
\begin{eqnarray*}
  \bu_{\dagger} &=& \left ( \bH^{-1\!} -
  \frac{p_{K+1\,}}{\deno} \bH^{-1} \bx_{K+1\,}
  \bx_{K+1}\tra \bH^{-1\!} \right )_{\!\!}
    \left ( \bs + p_{K+1\,} t_{K+1\,} \bx_{K+1} \right )
    \\[1mm] &=&
    \bu \,+\, p_{K+1} \left ( t_{K+1\,} 
    \,-\, \frac{\bx_{K+1\,}\tra \bu}{\deno}
\right .
    \\[1mm]  && \left . 
    \,-\,~ \frac{p_{K+1\,} \bx_{K+1\,}\tra \bH^{-1} \bx_{K+1}}{\deno}_{\,} 
     t_{K+1} \right ) \bH^{-1} \bx_{K+1}
    \\[1mm] &=& 
    \bu \,-\, \frac{p_{K+1\,} \delta_{K+1}}{\deno}_{\,} \bH^{-1} \bx_{K+1} \,,
\end{eqnarray*}
where $\delta_{K+1} = \bx_{K+1\,}\tra\bu - t_{K+1}$ 
is the discrepancy for the new auxiliary variable
evaluated with the original calibrated weights $\bu$.}
Hence the change in the estimate, 
$\Delta\htheta = \htheta_{\dagger} - \htheta =
\bu_{\dagger}\tra \by - \bu\tra\by$, is 
$$
\Delta\htheta \,=\, - \frac{p_{K+1\,} \delta_{K+1}}{\deno}
    \bx_{K+1\,}\tra \bH^{-1} \by \,.
$$
The denominator is greater than 1, so
$$
\left | \Delta \htheta_{\,} \right | \,<\, p_{K+1} \left |_{\,} \delta_{K+1\,}
\bx_{K+1\,}\tra \bH^{-1} \by_{\,} \right | \,.
$$
We assume upper bounds on $p_{K+1}$ and $|_{\,}t_{K+1\,}|$
and find the variable $\bx_{K+1}$ for which the upper bound
on $|_{\,}\Delta\htheta_{\,}|$ attains its maximum subject to the condition
that $\bx_{K+1}$ is standardised and orthogonal to $\bX$; that is,
$\bx_{K+1\,}\tra\bx_{K+1} = n$ and $\bX\tra \bx_{K+1} = \bzer_{K+1\,}$.
In a separate maximisation we impose the weaker condition
of standardisation only, that is,
  $\bx_{K+1}\tra \bone = 0$ and $\bx_{K+1\,}\tra\bx_{K+1} = n$,
so that $\bx_{K+1}$ may be correlated with variables in $\bX$.
Such a variable $\bx_{K+1}$ may duplicate some information contained in $\bX$.

A solution without the orthogonality constraint, denoted by $\bx^{(0)}_{K+1\,}$, 
assesses the sensitivity to the compendium of additional auxiliary information
and $\bp$ because the switch from $\bX$ to $\bX_{\dagger}$
may affect some priorities in $\bp$;
see Supplementary Materials A for details.
A solution orthogonal to $\bX$ assesses the sensitivity of $\htheta$
solely to additional auxiliary information
because the switch from $\bX$ to $\bX_{\dagger}$ 
leaves $\bp$ intact.
If the solutions $\bx_{K+1}$ and $\bx^{(0)}_{K+1}$ are highly correlated,
so that even $\bx_{K+1}^{(0)}$ is nearly orthogonal to $\bX$,
then $\htheta$ is sensitive mainly to the new information
and concerns about the specification of $\bp$ can be largely allayed.
In contrast, when $\bx_{K+1}^{(0)}$ is close to or belongs to the space
generated by $\bx_{1\,}, \ldots, \bx_{K\,}$,
and therefore nearly orthogonal to $\bx_{K+1\,}$,
the two solutions are nearly orthogonal.  
In that case $\htheta$ is sensitive mainly to $\bp$.
Therefore the sample correlation of the solutions
$\bx_{K+1}$ and $\bx_{K+1}^{(0)}$ has some diagnostic value.

We further assume that $K \ll n$ and $\bX$ has full column rank $K+1$.
The latter technical condition entails no loss of generality
because the redundant columns of $\bX$ can be removed
and their priorities apportioned to the retained columns of $\bX$. 
The extreme-case variable $\bx_{K+1}$ is found by
the method of Lagrange multipliers, searching for the extremes
of the function
\begin{equation} \label{Lagr}
f(\bx_{K+1\,}; \blambda_{1\,}, \lambda_2) \,=\,
- \delta_{K+1\,} \bx_{K+1\,}\tra \bH^{-1} \by
+ \bx_{K+1\,}\tra \bX_{\,} \blambda_{1\,}
- \frac{1}{2}_{\,}\lambda_2 \left ( \bx_{K+1\,}\tra\bx_{K+1} - n \right )
\,. ~~~
\end{equation}
Denote $\bQ = \bu_{\,} \by\tra \bH^{-1} + \bH^{-1} \by_{\,} \bu\tra$;
this is a symmetric $n \times n$ matrix of rank no greater than 2.
The identity
$$
  \frac{\partial f}{\partial \bx_{K+1}} \,=\, - \bQ \bx_{K+1} +
  t_{K+1\,} \bH^{-1} \by + \bX_{\,} \blambda_1 - \lambda_{2\,} \bx_{K+1} 
$$ 
implies that the solution has to satisfy the equation
\begin{equation} \label{Difr}
\bx_{K+1} \,=\, \left ( \lambda_{2\,} \bI + \bQ \right )^{-}
\left ( t_{K+1\,} \bH^{-1} \by + \bX_{\,} \blambda_1 \right ) \,,
\end{equation}
where $\bA^-$ denotes a generalised inverse of square matrix $\bA$.
We set aside the two values of $\lambda_2$ for which $\lambda_{2\,} \bI + \bQ$
is singular and explore them separately.
By pre-multiplying equation \refb{Difr} by $\bX \tra$,
which eliminates its left-hand side,
we obtain
$$
\blambda_1 \,=\,
- t_{K+1} \left \{ \bX \tra \bRQ
\bX \right \}^{-1} \bX \tra \bRQ \bH^{-1} \by \,.
$$
Its substitution to equation \refb{Difr} yields the solution 
\begin{equation} \label{XK1}
  \bx_{K+1} \,=\, t_{K+1\,} \bZ \bH^{-1} \by \,,
\end{equation}
where
$$
\bZ \,=\,   \left [ \bI  - \bRQ \bX \left \{
\bX \tra \bRQ \bX \right \} ^{-1} \bX \tra \right ] \bRQ
$$
and $\lambda_2$ is such that $\bx_{K+1\,}\tra\bx_{K+1} = n$.
The two values of $\lambda_2$ equal to the negatives of the
non-zero eigenvalues of $\bQ$,
as well as $\lambda_2 = 0$, have to be explored separately.
Assuming that $\bc = \bH^{-1}\by$ is not a scalar multiple of $\bu$,
  the two eigenvalues are
  $\bc\tra\bu \pm \sqrt{\bu\tra\bu {~} \bc\tra\bc}$;
  the eigenvectors are scalar multiples
  of $\bc\sqrt{\bu\tra\bu} \pm \bu \sqrt{\bc\tra\bc}$.
For the alteration of the estimate $\htheta$
with the solution $\bx_{K+1}$ we have the approximation
\begin{equation} \label{Apx1}
\Delta\htheta \,\doteq\, - p_{K+1\,} t_{K+1\,}^2
\left ( \bu\tra \bZ \bc - 1 \right ) \bc\tra \bZ \bc  \,.
\end{equation}
Although $p_{K+1}$  and $t^2_{K+1}$ are multiplicative factors
in this expression, $\bZ$ involves $t_{K+1}$ through the Lagrange
factors $\blambda_1$ and $\lambda_{2\,}$.  So, $\Delta\htheta$
is proportional to $p_{K+1\,}$ but its dependence on $t^2_{K+1}$
is not linear.
Note that $\bZ\bu$ and $\bZ\bc$
have to be evaluated anew for each value of $t_{K+1}$
but,
for a given $\by$,
$\bc$ has to be evaluated only once.

If instead of orthogonality we insist only on standardisation,
by replacing the Lagrange term $\bx_{K+1}\tra \bX \blambda_1$
in equation \refb{Lagr} with $\lambda_{1\,} \bone \tra \bx_{K+1}$ 
we obtain the solution $\bx^{(0)}_{K+1} = t_{K+1\,} \bZ^{(0)\,}\bc$
and the approximation
\begin{equation} \label{Apx2}
\Delta \htheta^{(0)} \,\doteq\, -p_{K+1\,} t^2_{K+1} \left (
\bu\tra\bZ^{(0)\,} \bc - 1 \right ) \bc\tra \bZ^{(0)\,} \bc \,,
\end{equation}
where
$$
\bZ^{(0)} \,=\, \left \{ \bI - \frac{1}{\bone \tra \bRQ \bone}
  \bRQ \bone \bone \tra \right \} \bRQ \,.
$$
The approximations in equations \refb{Apx1} and \refb{Apx2} 
  do not involve $\bx_{K+1}$ or $\bx_{K+1\,}^{(0)}$ directly.
  Of course, $\bx_{K+1}$ and $\Delta \htheta$ (or $\Delta \htheta^{(0)}$)
  have the term $\bZ\bc$ (or $\bZ^{(0)\,}\bc$) in common.
  The approximations
are based on a specific value $t_{K+1\,}$, and so their counterparts
for $t_{K+1} \in (-T, T)$ for a given $T > 0$ are found as the maxima
of \refb{Apx1} or \refb{Apx2} over a fine grid of values of $t_{K+1}$
in $(0, T)$, since the problem is symmetric in $t_{K+1\,}$.
  
In the evaluation of $\bx_{K+1}$ in equation \refb{XK1}
we have to form neither the matrix $\bH$
nor its inverse; see Longford (2024), Section 2.1.
Numerical inversion of $\bR = \lambda_{2\,} \bI + \bQ$ is avoided similarly.
In both cases we take advantage of the form of $\bH$ (and $\bR$)
as the sum of a matrix that is easy to invert, a scalar multiple of $\bI$,
and a matrix of rank much lower than $n$.
We also have the expression
$$
\bR^{-1} \,=\, \frac{1}{\lambda_2}_{\,} \bI 
+ A_{\,}\bc_{\,}\bu\tra + B_{\,}\bu_{\,}\bc\tra
- C_{\,} \bc_{\,}\bc\tra - D_{\,} \bu_{\,} \bu\tra ,
$$
where 
  $A = \bu\tra\bu_{\,} (\lambda_2 + \bc\tra\bu) / E$,
  $B = \bc\tra\bc_{\,} (\lambda_2 + \bc\tra\bu) / E$, 
  $C = D = \bc\tra\bc_{\,} \bu\tra\bu / E$ 
  and $E = \lambda_{2\,} \{ \bc\tra\bc_{\,} \bu\tra\bu -
( \lambda_2 + \bc\tra\bu )^2 \}$.

The optimal vector $\bx_{K+1}$ is found by the Newton method or any other
method of line search applied to $\lambda_{2\,}$. 
It is practical to plot the function $\bx_{K+1\,}\tra \bx_{K+1} - n$ first
to find a suitable
pair of initial values of $\lambda_2$ for the iterations.

It is not useful to impose an upper bound on $|_{\,}\delta_{K+1\,}|$
instead of $|_{\,}t_{K+1\,}|$ 
because a solution may be obtained for which both 
$\bx_{K+1}\tra \bu_{\dagger}$ and $t_{K+1}$ have large absolute 
values and yet their difference, the discrepancy $\delta_{K+1,\dagger\,},$
is small.
However, such a value of $t_{K+1}$ may be implausible.
Supplementary Materials B describe the algorithm on which this conclusion
can be verified.
\section{Examples and simulations} \label{Ssim}

We generated a population of size $N = 120\,000$
with an outcome variable $y$ and a set of $K = 12$
auxiliary variables, six each in the roles of high-priority (important)
and low-priority (unimportant) variables.
The outcome variable $y$ is generated as a linear combination
of the high-priority variables and an additive random term
and the probabilities of selection are generated
from three of these high-priority variables.
  Poisson sampling design with expected sample size 1000
  is used.
See Supplementary Materials C (SM-C) for details, 
including histograms of the probabilities and outcomes in SM-C Figure 1.
The priority parameters are set to 3.0 for the intercept ($\bx_0$),
0.1 for each of the six important variables in $\bX$
and 0.01 for each of the six other (unimportant) variables.
The priority for small alteration of the weights, $R$, is set to 0.5.  

For illustration, we first use a single population, ${\cal{P}}$,
draw from it a single sample ${\cal{S}}$, apply calibration
and then conduct the sensitivity analysis for a new auxiliary variable
$\bx_{K+1\,}$.
We set its largest plausible priority $p_{K+1}$ to 0.1
and the largest plausible deviation to $t_{K+1} = 5000$.
After describing the results for this setting, 
we alter one of its elements at a time and explore
how the results are changed or how they vary as a function
of the selected parameter.

The target, the population total of $y$ is $\theta = 1236.6$.  
Its Horvitz-Thompson (HT, or pre-calibration) estimate is
$\bw\tra\by = 1317.6$, with error 80.9.
The calibration estimate is equal to $\bu\tra\by = 1241.9$,
with error 5.3.
If the population total of $y$ were not known, the quality of the calibration
would be assessed by the magnitude of the absolute discrepancies
$|_{\,}\delta_{k\,}|$, $k=0, 1, \ldots, K$.
They are the absolute errors in estimating $t_k$ based on calibrated
weights $\bu$.
While the absolute errors in estimating $t_k$ based on $\bw$
(prior to calibration) are in the range 1250\,--\,7900,
they are reduced by calibration, using $\bu$, to 1\,--\,641,
and the largest of them for the important auxiliary variables
is only 180.  The error for the intercept is $\delta_0 = 1.0$.
See SM-C Table 1.

We found that, as a function of the Lagrange multiplier $\lambda_{2\,}$, 
$\log(\bx_{K+1}\tra \bx_{K+1})$ has much less curvature
than $\bx_{K+1}\tra \bx_{K+1\,}$.  
It is therefore easier to solve the equation $\bx_{K+1}\tra \bx_{K+1} = n$
by log-transforming its two sides --- the domain of convergence is wider
and fewer iterations are required.  See SM-C Figure 2.

  The upper bound for the change $|_{\,}\Delta\htheta_{\,}|$
  in the calibration estimate of $\theta$ is 0.33
  with the orthogonality constraints, 
  $\bX\tra\bx_{K+1} = \bzer$, and $2.60$ without them.
We emphasize that these bounds are specific to the sample and the outcome  
variable, as well as to $p_{K+1}$ and $t_{K+1\,}$.
In the analysis stage the sensitivity for the realised sample is of interest.
Selecting a sampling design, subject to the various constraints on resources,
practicality and desired precision of key estimators,
in which an estimator $\htheta$ would be least sensitive
to new auxiliary information is an unsolved problem.

Compared with the error in estimation, 5.31,
the bounds for $\Delta\htheta$ are small.
We conclude that the estimates are insensitive to new auxiliary information
and somewhat more to the priorities $\bp$.
In practice, without $\theta$ known, the analyst
might relate
the bounds on $\Delta\htheta$ and $\Delta\htheta^{(0)}$
to the estimated standard error of $\htheta$, 
$s(\htheta)$, even though the sensitivity and sampling variation assess
different aspects of uncertainty about $\theta$.
Whilst $s(\htheta)$ is an estimate of a design-related quantity,
the sensitivity bounds entail no reference to replicate samples.
The standard error of the HT estimator, evaluated from ${\cal{P}}$,
is 40.7, and its estimate, based on ${\cal{S}}$, is 40.3,
many times greater than $\Delta\htheta^{(0)}$. 

\begin{figure}

\includegraphics[width=145mm]{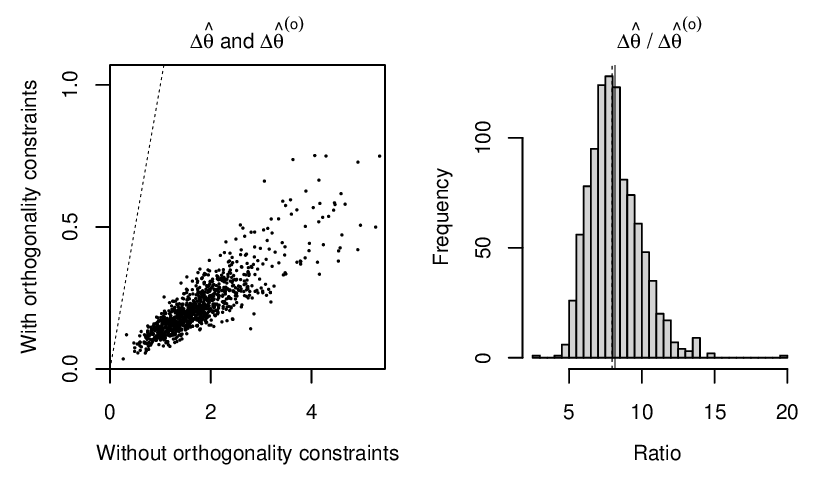}
  
\caption{The simulated estimates $\Delta\htheta$ and $\Delta\htheta^{(0)}$
  (left) and their ratios (right).
  The line drawn by dashes on the left indicates the identity.}
\label{FSM3}
\end{figure}

We conducted simulations with 1000 replications of the sampling
and calibration processes and evaluations
of $\Delta \htheta$ and $\Delta \htheta^{(0)}$, to explore
to what extent the results obtained for the original sample ${\cal{S}}$
are a property of the sample drawn versus the processes involved.
The population ${\cal{P}}$ was fixed throughout.
Figure \ref{FSM3} displays a plot of the replicate values
of $\Delta \htheta$ and $\Delta \htheta^{(0)}$.  It shows
that the pairs of estimates are highly correlated (0.862).
The empirical means (standard deviations) of $\Delta \htheta$
and $\Delta \htheta^{(0)}$ are 0.226 (0.102) and 1.791 (0.768), respectively.
The median of the ratio $\Delta \htheta^{(0)} / \Delta \htheta$ is 7.96
(interquantile range 6.94\,--\,9.18).
The sample correlations of $\bx_{K+1}$ and $\bx_{K+1}^{(0)}$
have mean 0.241 and standard deviation 0.034.
In summary, sensitivity to a new auxiliary variable in the studied
setting is governed mainly by uncertainty about the priorities $\bp$,
but even that is very small. 

These simulations also confirm that the calibration process operates
as intended.  
The replicate discrepancies $\delta_0$ are tightly concentrated
around zero, with standard deviation 0.67, reflecting the high priority
($p_0 = 3.0$).  The standard deviations of the discrepancies $\delta_k$
for the important variables are around 40 and for unimportant variables
around 375.  Without calibration, they are around 4000.
Table \ref{TSM1} presents the details.
For completeness, the empirical standard errors of the HT and calibration
estimators are 39.9 and 2.54, respectively.  

\begin{table}[t!]
  \caption{Standard deviations of the replicate discrepancies
    $\delta_{k,\dagger\,}, k=1, \ldots, K=12$; based on 1000 replications.
    The standard deviation of $\delta_{0,\dagger}$ is 0.67.  The empirical
  means of all $\delta_{k,\dagger}$ are close to zero.}
    \label{TSM1}
\begin{center} {\tabcolsep 9.09pt \begin{tabular}{lcrrrrrr}
\\[-7.5mm] \hline \noalign{\vspace{1mm}}
Variables & Priority & 1~~ & 2~~ & 3~~ & 4~~ & 5~~ & 6~~
\\[1mm] \hline \noalign{\vspace{1mm}}
Important & 0.10 & 41.8 & 37.1 & 39.7 & 39.2 & 40.3 & 38.5 \\[1mm]
Unimportant & 0.01 & 375.6 & 364.5 & 367.3 & 381.6 & 383.1 & 365.7 
\\[1mm] \hline \noalign{\vspace{1mm}}  
    \end{tabular}} \end{center}
  \end{table}

As a function of $t_{K+1\,}$, the largest plausible value of $\Delta \htheta$
increases up to about 5800, is constant in the range 5800\,--\,13\,800,
and then increases sharply.
A similar anomaly occurs for $\Delta \htheta^{(0)}$ for much greater values
of $t_{K+1\,}$; see SM-C Figure 4. 
These anomalies recur when the priorities $\bp$ are specified inappropriately
(low for important and high for unimportant variables) and when,
in a different population, $\by$ depends on $\bX$ more weakly;
see SM-C Figures 5\,--\,7.
Weak dependence is associated with lower sensitivity and inappropriate
priorities are `punished' by higher sensitivity.

Dependence on $R$ is explored in SM-C Figure 8.
Lower $R$ is associated with lower sensitivity; with increasing $R$
the anomaly of constant $\Delta \htheta$ moves to higher values of $t_{K+1\,}$.

Sensitivity increases with sample size for both $\Delta \htheta^{(0)}$
and $\Delta \htheta$, presumably only for sample sizes much smaller
than the population size because as sample size $n$ approaches
$N$ the estimation error diminishes,
and so do $\Delta \htheta^{(0)}$ and $\Delta \htheta$.
See SM-C Figure 9 for details.  Less congenial settings,
in which $\by$ depends (strongly) on some unobserved variables,
  are discussed at the end of this section.

SM-C Figure 10 explores the dependence of sensitivity and estimation
error on the priorities $\bp$, using a fixed sample.
Higher priority is associated with smaller estimation error.
Sensitivity increases with $p_0$ but decreases with $p_{k\,}$,
$k=1, \ldots, K$, approaching in every case an asymptote.
The estimation errors, $\htheta -\theta$, also approach an asymptote
and, for $p_0$ in a wide range ($p_0 > 1$), $\htheta$ is in an extremely
narrow range, (1.498, 1.499).  Thus, sensitivity and efficiency
are salient considerations only for small values of $p_{k\,}$.

We emphasise that these findings relate to a particular setting,
a population in particular, and they may not apply in other settings.
However, they provide a template for exploration of sensitivity
in other populations, sampling designs, and auxiliary information.

The extreme plausible variables $\bx_{K+1}$ and $\bx_{K+1}^{(0)}$
are sample-specific, are the result of a search,
in a given linear space, and are informed by the outcome.  
Therefore neither variable can be regarded as a discovered correlate
of the outcome $y$;
their inclusion in a model for $y$ in terms of the auxiliary variables
would be inappropriate.  The improved `explanation' and reduction
of the residual sum of squares by either of them are fallacious
because of potentially gross capitalisation on chance.
Suppose the outcomes $\by$ are generated by a linear model
with variables in $\bX_+ = (\bX, \bx_+)$.
If variable $\bx_+$ is omitted from calibration, then it is
nearly recovered by re-calibration, as $\bx_{K+1\,}$, 
only when the residual variance $\var(\by_{\,}|_{\,}\bX_+)$ is very small.
This issue is explored by simulations in Supplementary Materials D.

\section{Discussion} \label{Disc}

Credible evidence of completeness of the auxiliary information 
for calibration in a survey can rarely be obtained.
It is therefore pertinent to ask how much the calibration estimate
$\htheta$ of a population total would be altered
if the  auxiliary variables currently used were supplemented
by another variable --- how sensitive the calibration estimate
is to new auxiliary information.
A constructive answer has to impose some limits
on such a new variable to avoid implausible settings,
such as a variable with extremely large discrepancy
or an unrealistically important variable, 
such as a strong correlate of the outcome.
For these limits we propose, with relaxed calibration,
the largest plausible priority $p_{K+1}$
and the largest plausible population total $t_{K+1}$
of the new auxiliary variable after its standardisation.
Setting these limits may appear to be a challenging task,
exposing the analyst to (unfair) criticism when the context
of the study is better understood at some point in the future.
However, the examples in Section \ref{Ssim} suggest
that the exact choices of $p_{K+1}$ and $t_{K+1}$ are not crucial
for the assessment of sensitivity.
The method developed in this article, 
constructs the sample version of an auxiliary variable
for which $\htheta$ would be altered most, within the set limits, 
$p_{K+1}$ and $t_{K+1\,}$.

Relaxed calibration, as well as its add-on presented here,
are entirely design-based but they do not preclude involvement of models,
in an auxiliary role in particular.
For instance, there is a potential for the bounds found by our method
to be sharpened by incorporating prior (expert) information,
possibly with model-based origins,
about the nature of the missing auxiliary variable(s).
Such information is likely to be accumulated in abundance
in regularly conducted surveys with a long history, especially
when the surveyed population evolves without any sudden changes
and is studied by multiple parties.

The extreme plausible variables $\bx_{K+1}$ and $\bx_{K+1}^{(0)}$
constructed by our method are specific to the sample as well
as the outcome variable.  When there are several outcomes
we obtain multiple pairs of such extreme variables.
Although none of them can be used as a discovered auxiliary variable,
a secondary analysis of these variables may reveal some features
of the missing auxiliary information.
For a set of outcomes that are not strongly correlated,
the collection of the upper bounds $|_{\,}\Delta\htheta_{\,}|$
may be too pessimistic because each of these bounds is associated
with different new information.
We suggest developing these ideas for future research.

  Apart from the plausible change of the estimate
  $\htheta$ by re-calibration, the plausible change in the
  sampling variance of the re-calibrated estimator may also be
  of interest, and in particular that it would not be increased.
  The would-be estimator of the sampling variance based on
  $(\bX, \bx_{K+1})$ or $(\bX, \bx^{(0)}_{K+1})$ is biased
  because the new auxiliary variable is treated
  as fixed (without sampling variation) and 
  as if it were constructed without using the outcome.
  Moreover, the sampling variance may be reduced more
  by using a plausible new auxiliary variable different
  from $\bx_{K+1}$ and $\bx^{(0)}_{K+1\,}$.

Drawing a parallel with design sensitivity in causal inference, 
(Rosenbaum, 2004; Howard and Pimentel, 2021),
there may be some scope for adjusting the design of a survey
so that the calibration of the subsequently collected data 
using the available auxiliary information would be less sensitive
to its incompleteness.
From the derivations of the largest plausible change $\Delta\htheta$
it is not transparent how the extreme-case new variable, $\bx_{K+1}$
or $\bx_{K+1\,}^{(0)}$, is related to any features of the sampling design
and the available auxiliary information.
These issues and, more generally, a theoretical support
for the empirical findings in the examples and simulations, 
are challenges for further research.

\subsubsection*{Acknowledgements and supplementary materials}

Reviewer's insightful comments and suggestions have contributed
to substantial improvements on the originally submitted manuscript.

Supplementary materials for this article comprise four sections.
Section A discusses invariance with respect to linear transformations, 
Section B describes an alternative algorithm for sensitivity analysis,
Section C gives details of examples and simulations.
Section D addresses, by simulations, the issue of interpreting
the extreme plausible variable $\bx_{K+1}$ 
as a discovered covariate for the outcome
and Section E contains information about software.

\section*{References}

\parindent 0pt \parskip 1.5mm

Chambers, R., Skinner, C.J., and Wang, S. (1999).
Intelligent calibration.
{\em Bulletin of the International Statistical Institute:
  52nd Session Proceedings}, pp.\ 321--324.
International Statistical Institute, the Hague, the Netherlands.

Deming, W.E., and Stephan, F.F. (1940).
On a least squares adjustment of sampled frequency table
when the expected marginal totals are known.
{\em Annals of Mathematical Statistics} {\bf 11}, 427--444.

Devaud, D., and Till\'e, Y. (2019).
Deville and S\"arndal's calibration: revisiting a 25-year-old
successful optimization problem. {\em Test} {\bf 28}, 1033--1065.

Deville, J.-C., and S\"arndal, C.-E. (1992).
Calibration estimators in survey sampling.
{\em Journal of American Statistical Association} {\bf 87}, 1013--1020.

Haziza, D., and Beaumont, J.-F. (2017).
Construction of weights in surveys:  A review.
{\em Statistical Science} {\bf 32}, 206--226. 

Howard, S.R., and Pimentel, S.D. (2021).
The uniform general signed rank test and its design sensitivity.
{\em Biometrika} {\bf 108}, 381--396.

Kwon, Y., Kim, J.K., and Qiu, Y. (2025).
Debiased calibration estimation using generalized entropy
in survey sampling.
{\em Journal of the American Statistical Association} {\bf 120};
to appear.  Published online on 30th September 2025;
{\tt https://doi.org/10.1080/01621459.2025.2537452}.

Longford, N.T. (2024).
Relaxed calibration of survey weights.
{\em Survey Methodology} {\bf 50}, 261--285.

Potthoff, R.F., Woodbury, M.A., and Manton, K.G. (1992).
`Equivalent sample size' and `equivalent degrees of freedom'
refinements for inference using survey weights under superpopulation models.
{\em Journal of the American Statistical Association} {\bf 87}, 383--396.

Rosenbaum, P.R. (1995).
{\em Observational Studies.}
Springer-Verlag, New York.

Rosenbaum, P.R. (2004).
Design sensitivity in observational studies.
{\em Biometrika} {\bf 91}, 153--164.

\clearpage

\begin{center}\section*{Supplementary materials}\end{center}

\subsection*{A. Linear invariance} 

Relaxed calibration is invariant with respect to linear transformations
of the auxiliary data $\bX$ in the following sense.
Let $\bE$ be a $(K+1) \times (K+1)$ nonsingular matrix and denote
$\bX_{\rm E} = \bX \bE$, $\bt_{\rm E} = \bE \bt$ and $\bP_{\rm E} = \bE^{-1} \bP \bE^{-1}$.
Then for $\bH$ and $\bs$ defined in Section 2,
$\bH^{-1} \bs = \bH^{-1}_{\rm E\,} \bs_{\rm E\,}$, where
$\bH_{\rm E}$ and $\bs_{\rm E}$ are the respective counterparts
of $\bH$ and $\bs$ in equation (1) with $\bX$, $\bt$ and $\bP$
replaced by $\bX_{\rm E\,}$, $\bt_{\rm E}$ and $\bP_{\rm E\,}$, respectively.
Note that $\bP_{\rm E}$ is positive definite but, unlike $\bP$,
not necessarily diagonal.

This invariance implies that the auxiliary information is fully specified
by a linear subspace of ${\cal{R}}^n$, and $\bX$ is well represented
by any basis of this subspace.
This characterisation reveals an ambiguity about the priorities $\bp$.
By way of an example, suppose $K\ge 2$ and the auxiliary variable $\bx_2$
is $\bx_2 = a \bx_1 + \be$, where $a \in (-1,1)$ is a scalar and $\be$
is such that $\bx_1\tra\be = 0$;
$\bx_1$ and $\bx_2$ have respective priorities $p_1$ and $p_{2\,}$.
Then $\bx_2$ can be replaced in $\bX$ by $\be / \sqrt{1 - a^2}$ but
$p_1$ has to be replaced by $(1+a)^2p_1$ and $p_2$ by $(1 - a^2)_{\,}p_{2\,}$.
In summary, the priority of an auxiliary variable ($\bx_1$)
has to take into account the other variables in $\bX$.
Orthogonalisation of $\bX$ might resolve this issue
but such a transformation is not unique and the orthogonal variables
would be difficult to interpret and their priorities specified. 

\subsection*{B.  Stability of $\htheta$ with a constraint on $\delta_{K+1}$}

Suppose an upper bound is specified for $|_{\,}\delta_{K+1\,}|$
  instead of $|_{\,}t_{K+1\,}|$.
Then for $f$ in equation (2), with the constraint $\bX\tra\bx_{K+1} = \bzer$, 
we have the identity
$$
\frac{\partial f}{\partial \bx_{K+1}} \,=\, - \delta_{K+1\,}\bH^{-1} \by
+ \bX_{\,} \blambda_1 - \lambda_{2\,} \bx_{K+1} \,,
$$
from which we obtain the root
$$
\blambda_ 1 \,=\, \frac{1}{n}_{\,} \delta_{K+1} \bX\tra\bH^{-1}\by 
$$
and the solution
$$
\bx_{K+1} \,=\, \frac{\delta_{K+1}}{\lambda_2} \left ( \frac{1}{n}_{\,} 
      \bX_{\,} \bX\tra - \bI_n \right ) \bH^{-1}\by \,,
$$      
where $\lambda_2$ is a normalizing constant for which $\bx_{K+1\,}\tra \bx_{K+1} = n$.
The largest plausible change of the estimate is $\Delta \htheta \,\doteq\, 
p_{K+1\,}|_{\,} \delta_{K+1\,}|_{\,} \by\tra\bH^{-1} ( n^{-1} \bX_{\,} \bX\tra - \bI_n )_{\,}
  \bH^{-1}\by$.

If instead of the constraint $\bX_{\,}\tra\bx_{K+1} = \bzer$
we insist only on $\bone_n\tra\bx_{K+1} = 0$, we obtain the solution 
$$
\bx_{K+1}^{(0)} \,=\, \frac{\delta_{K+1}}{\lambda_2} \left ( \frac{1}{n}_{\,} 
      \bone_{n\,} \bone_n\tra - \bI_n \right ) \bH^{-1}\by 
$$      
with $\Delta \htheta^{(0)} \,\doteq\, 
p_{K+1\,}|_{\,} \delta_{K+1\,}|_{\,} \by\tra\bH^{-1} ( n^{-1} \bone_{n\,} \bone_n\tra -
\bI_n )_{\,} \bH^{-1}\by$.

\subsection*{C.  Illustrations and simulations}  \label{IlSim}

We construct a population ${\cal{P}}$ of $N=120\,000$ units
and retain it for all the illustrations and simulations in this section,
except when stated otherwise.
In this population, we generate 12 mutually independent auxiliary variables,
two each drawn from the standard normal distribution,
chi-squared with four degrees of freedom, log-normal (exponential
of the standard normal), binary with probability 0.12, Poisson
with mean 2.5 and gamma with shape 1.0 and rate 5.0 (mean 0.2 and
standard deviation 0.2).  In each pair of these variables we designate
one as important and the other as unimportant. 
The outcome variable is generated as $\by = \bX \bbeta + \bepsi$,
where the elements of $\bbeta$ are equal to 1.0 for the intercept
and each important variable and 0.1 for every unimportant variable;
$\bepsi$ is generated as a random sample, unrelated to $\bX$,
from the centred normal distribution with standard deviation 0.4.
The variable is then truncated to be within the range $[0, 25]$.
In the realisation used, four negative values were re-defined
as zero and 649 (0.5\%) were truncated at 25.0.

\begin{figure}

\includegraphics[width=140mm]{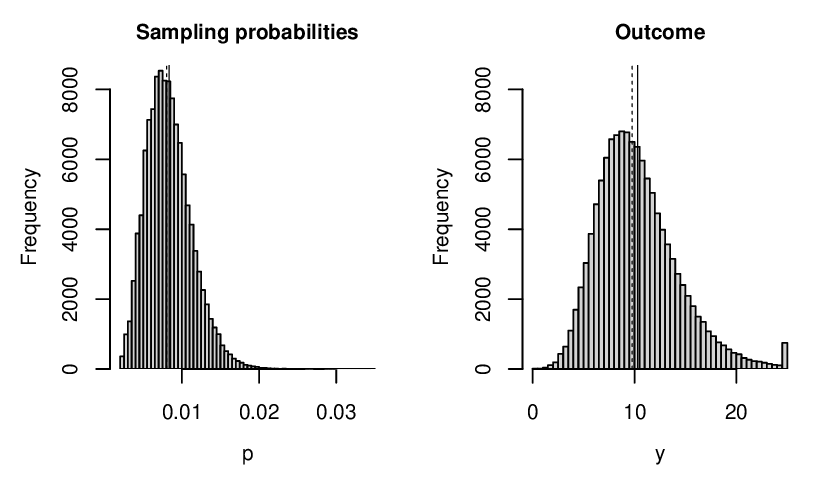}
  
\caption{Histograms of the sampling probabilities (left) and outcomes (right)
    in population ${\cal{P}}$.  
    The mean is marked by solid vertical line and median by dashes.}
  \label{FSM1}
\end{figure}

The sampling probabilities are generated as
$\bpi = c(1 + \bX \bgamma)$ where the elements of $\bgamma$ are equal
to zero for all variables in $\bX$ except for the three important
auxiliary variables:  the log-normal (coefficient 0.35), Poisson (0.7)
and gamma (0.4).  The factor $c$ is set so as to make the expected sample
size $\mE(n)$  equal to 1000.
The inclusions in the sample are mutually independent.
Figure \ref{FSM1} presents histograms of the sampling probabilities
and the outcomes.  
The target of estimation, the population total of $y$,
is $\by \tra \bone_{N\,} / N = 1\,236\,613$.
In all reporting we divide in by 1000 and round it
as well as all its estimates to one decimal place
($\theta = 1236.6$) or more when comparing smaller values.

The sample used below, denoted by ${\cal{S}}$, has size $n=1069$.
The Horvitz-Thompson (HT) estimate is equal to 1317.6, so its error is 80.9.
The calibration estimate is equal to 1241.9, with error 5.3.

Without any reference to the outcome variable, we assess the success
of calibration by comparing the errors in estimation of the sample
size $N$ and population totals of the auxiliary variables, committed
by the HT and calibration estimates.  Table \ref{TSM1a} displays
the estimates and errors.  The errors are reduced by calibration
substantially.  For four of the important variables they are smaller
in absolute value than for their unimportant counterparts.
With one exception, the absolute error of the calibration estimate
is over ten times smaller than for the corresponding HT estimate.  
The estimate of the population total of $y$ is also much closer to the target
after calibration, 1241.9 (error 5.3), than without it, 1317.6 (80.9).

\begin{table}[t]
  \caption{Horvitz-Thompson and calibration estimates and errors
    for the population totals of the auxiliary variables;
    sample ${\cal{S}}$ from population ${\cal{P}}$.  Important variables
    (printed with initial letter boldface) have priority 0.10
    and unimportant variables 0.01.}
  \label{TSM1a}
{\tabcolsep 6.8pt \begin{tabular}{lrrrrrrrr}
    \\[1mm]
    \hline \noalign{\vspace{1mm}}
    & Priority & \multicolumn{2}{c}{Estimate} && \multicolumn{1}{c}{Target}
    && \multicolumn{2}{c}{Error} \\[1mm]
    \cline{3-4} \cline{8-9} \noalign{\vspace{1mm}}
    Variable & \multicolumn{1}{c}{$p$} & HT~ & Calibration  && (Estimand) && HT~ & Calibration \\[1mm]
        \hline \noalign{\vspace{1mm}}
Intercept &   3.00~~~ &  126\,307   &  120\,001 &&  120\,000~\,~ &&    6307  &     1 \\[0.5mm]
{\bf N}ormal &      0.10~~~ &    --2553   &       969 &&      1007~\,~ &&  --3561  &  --38 \\[0.5mm]
Normal &      0.01~~~ &    --1604   &     --348 &&     --212~\,~ &&  --1392  & --135 \\[0.5mm]
$\bchi^2(4)$ & 0.10~~~ & --26\,807   & --28\,895 && --29\,049~\,~ &&    2242  &   154 \\[0.5mm]
$\chi^2(4)$ & 0.01~~~ &      2559   &     --118 &&     --199~\,~ &&    2758  &    81 \\[0.5mm]
{\bf L}og-normal  & 0.10~~~ &       111   &      1988 &&      1997~\,~ &&  --1886  &   --9 \\[0.5mm]
Log-normal  & 0.01~~~ &    --1284   &      5963 &&      6604~\,~ &&  --7888  & --641 \\[0.5mm]
{\bf B}inary  &     0.10~~~ &      1724   &    --4175 &&    --4222~\,~ &&    5946  &    47 \\[0.5mm]
Binary  &     0.01~~~ &     --345   &      6200 &&      6802~\,~ &&  --7148  & --602 \\[0.5mm]
{\bf P}oisson &     0.10~~~ & --32\,102   & --33\,176 && --33\,356~\,~ &&    1254  &   180 \\[0.5mm]
Poisson &     0.01~~~ &        99   &      1841 &&      1921~\,~ &&  --1823  &  --81 \\[0.5mm]
$\bGamma(1, 0.2)$ & 0.10~~~ & --2389 &    --8765 &&    --8839~\,~ &&    6450  &    74 \\[0.5mm]
$\Gamma(1, 0.2)$ & 0.01~~~ &   1025 &      3188 &&      3352~\,~ &&  --2327  & --164 \\[1.5mm]
{\em Outcome} & --- ~~~\, &     1242 & 1318 &&  1237~\,~ && 81 & 5 
\\[1mm]  \hline \noalign{\vspace{1mm}}
  \end{tabular}}
\end{table}

We set the parameters for analysing the sensitivity to a new auxiliary variable
to $p_{K+1} = 0.10$ and $t_{K+1} = 5000$.  So, we assume that
the new auxiliary variable could be as important as the six original
variables designated as `important'.
With the restriction to variables orthogonal to all the observed auxiliary
variables ($\bX\tra\bx_{K+1} = \bzer$),
we obtain the largest plausible change of the calibration estimate 0.33;
without this constraint the largest plausible change is 2.60,
nearly eight times larger.  Compared to the error in estimation, 5.3,
they are small.  We conclude that the estimates are quite stable. 

\begin{figure}

\includegraphics[width=140mm]{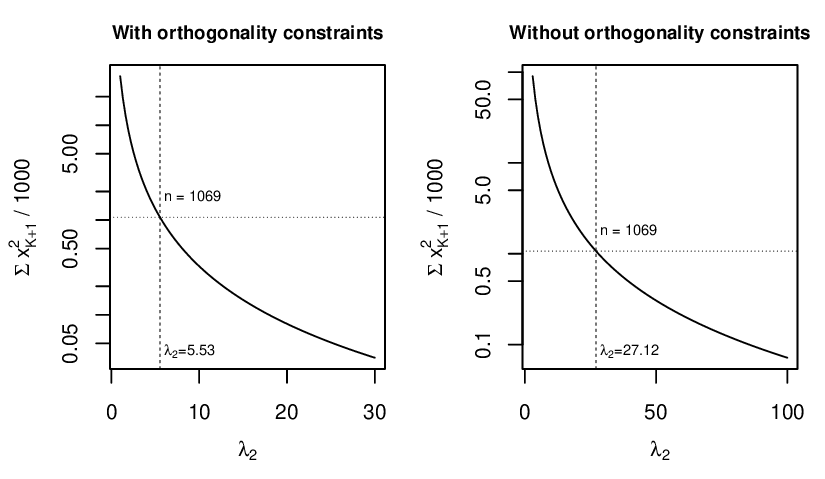}
  
  \caption{Solution of the equation $\bx_{K+1}\tra\bx_{K+1} = n$.}
  \label{FSM2}
\end{figure}

Figure \ref{FSM2} illustrates the solution of the equation
$\bx_{K+1}\tra\bx_{K+1} = n$
in the search of the Lagrange multiplier $\lambda_{2\,}$.
The vertical scale is multiplicative in both plots;
function $\bx_{K+1}\tra\bx_{K+1}$ is distinctly log-concave in both settings,
with the constraint $\bX\tra\bx_{K+1} = \bzer$ and without it.   Convergence is
achieved from a wider domain of initial values and is faster for the equation
$\log(\bx_{K+1}\tra\bx_{K+1}) = \log(n)$.

\subsection*{Replications of the sampling process}

We replicated the processes of sampling, calibration and sensitivity
analysis to explore to what extent the results of the sensitivity analysis
are a property of the population and sampling design on the one hand
and the realised sample on the other.  We used 1000 replications,
discarding replications for which the automated search for $\lambda_2$
by the Newton method failed.  In the 1000 replications, involving
2000 searches, only one replication had to be discarded.

The calibration process reflects the priorities $\bp$.  The standard
deviation of the replicate discrepancies $\delta_k$ is 0.65 for
the intercept, around 40 (37.1\,--\,41.7) for the important auxiliary
variables and around 375 (364.3\,--\,382.9) for the unimportant
variables.  Without calibration, the standard deviations
of the discrepancies $\bw\tra\bX - \bt$ are around 4000 (3644\,--\,4201);
see Table 1 of the main article. 

The average of the largest plausible changes in the estimate of $\theta$
that is, $\Delta\htheta$, is 0.226 (standard deviation 0.102)
with the constraint $\bX\tra\bx_{K+1} = 0$, and for 
$\Delta\htheta^{(0)}$, without the constraint, it is 1.791 (0.768).
The correlation of these  pairs of maxima is 0.862.
The correlations of the pairs of `new' variables $\bx_{K+1}$
and $\bx_{K+1}^{(0)}$ for the extreme plausible scenario
is in the range 0.15\,--\,0.36, with mean 0.242 and median 0.239.

\begin{figure}

\includegraphics[width=140mm]{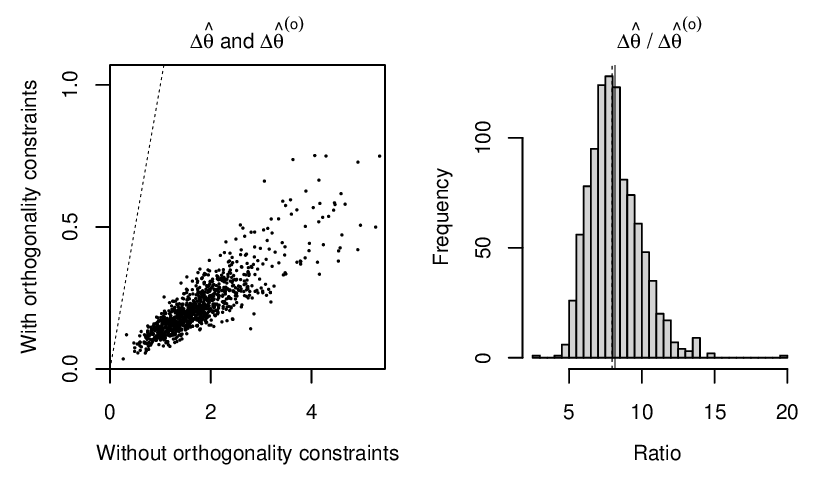}
  
\caption{Estimates $\Delta\htheta$ and $\Delta\htheta^{(0)}$
  (left) and their ratios (right).
    The dashes in the left-hand plot mark the identity line.}
  \label{FSM3a}
\end{figure}

Figure \ref{FSM3a} displays the plot of the pairs of replicate values
$\Delta\htheta$ and $\Delta\htheta^{(0)}$,
and the histogram of their ratios.
It shows that in the vast majority of samples, the largest change
$\Delta\htheta$ that would arise by a new auxiliary variable
correlated with the original variables is much greater
than by a variable orthogonal to $\bX$.
The median of the ratios of their values, marked by vertical dashes
in the histogram, is 7.96 (interquantile range 6.94\,--\,9.18),
slightly higher than their mean.
The two maxima are highly correlated (0.862); the replicate values
of $\Delta\htheta$ are much less dispersed than the values
of $\Delta\htheta^{(0)}$ (respective standard deviations
0.102 and 0.768).
Thus, sensitivity is caused mainly by uncertainty about the priorities.
We believe this owes to the sampling design, in which the probabilities
of inclusion are a deterministic function of the important auxiliary variables. 

\subsection*{Dependence on the expected sample size}

We repeated these simulations for expected sample sizes $\mE(n) = 500$
and 2000, with the same population and priority parameters.
The expected sample sizes were arranged by halving the original
inclusion probabilities for $\mE(n) = 500$ and doubling them 
for $\mE(n) = 2000$.  The standard deviations of the discrepancies
$\delta_{k\,}$, $k=0, \ldots, K$, behave in the predictable fashion ---
the variances decrease with sample size.
However, sensitivity to new auxiliary variable increases with sample size.
The average values of $\Delta\htheta$ are 0.176, 0.226 and 0.318
for the respective expected sample sizes 500, 1000 and 2000.
Their dispersion also increases; the standard deviations are
0.081, 0.102 and 0.128.  For $\Delta\htheta^{(0)}$
we have a similar pattern,
although with several times larger values.
The mean values are 1.483 (standard deviation 0.657), 1.791 (0.768)
and 2.395 (0.938) for the respective sample sizes 500, 1000 and 2000.  
The correlations of $\bx_{K+1}$ and $\bx_{K+1}^{(0)}$ are affected
by the sample size only slightly; their means are 0.236, 0.242 and 0.248
for $\mE(n) = 500, 1000$ and 2000, respectively.

\subsection*{Dependence on sensitivity parameters $p_{K+1}$ and $t_{K+1\,}$}
\begin{figure}


  \includegraphics[width=140mm]{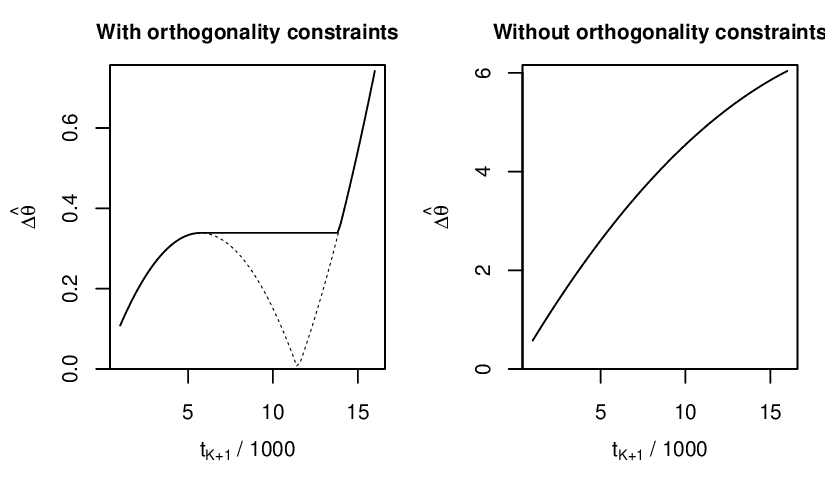}
  \caption{$\Delta\htheta$ and $\Delta\htheta^{(0)}$
    as functions of $t_{K+1\,}$.
    The values are drawn for the constraints $t_{K+1} \le T$
    by solid lines and $t_{K+1} = T$ by dashes.}

  \label{FSM4}
\end{figure}

Next, we study the dependence of $\Delta\htheta$
and $\Delta\htheta^{(0)}$
on the sensitivity parameters, with ${\cal{P}}$ and ${\cal{S}}$ fixed.
Throughout, all results are given for $p_{K+1} = 0.1$,
on par with an important auxiliary variable. 
First, $p_{K+1}$ is a multiplicative factor in both
$\Delta\htheta$ and $\Delta\htheta^{(0)}$.
The dependence on $t_{K+1}$ is more intricate because $t_{K+1}$
is involved in the normalising condition $\bx\tra_{K+1}\bx_{K+1} = n$.
Figure \ref{FSM4} displays the values
of $\Delta\htheta$ and $\Delta\htheta^{(0)}$ as functions of $t_{K+1\,}$.
The algorithm for evaluating them implements the constraint
that $t_{K+1}$ is equal to the specified value ($T$),
whereas we are interested in $|_{\,}t_{K+1\,}|$ being smaller than $T$.
The values for $t_{K+1} = \pm T$ are drawn by solid lines and for
$|_{\,}t_{K+1\,}| \le T$ by dashes.  They differ only when the former
is not increasing, as in the case in the left-hand panel,
for $\Delta\htheta$,
it is constant in the range $5800 < t_{K+1} < 13\,800$.
At $t_{K+1} \doteq 11\,400$, $\Delta\htheta(t_{K+1})$
attains its minimum,
close to zero, and then increases steeply.  This value of $t_{K+1}$
appears not to be related to the eigenvalues of
$\bQ = \bu \by\tra \bH^{-1} + \bH^{-1}\by \bu\tra$. 
For $\Delta\htheta^{(0)}$ such an anomaly occurs at much greater
values of $t_{K+1\,}$.

\subsection*{Inappropriate priorities and weaker dependence of $\by$ on $\bX$}

\begin{figure}


\includegraphics[width=140mm]{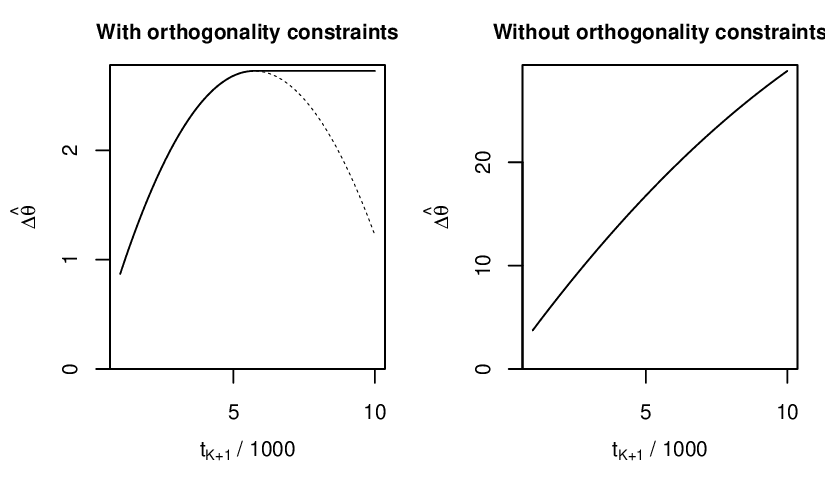}
  
\caption{Extreme plausible values of $\Delta\htheta$
  and $\Delta\htheta^{(0)}$ as functions
    of $t_{K+1}$ for priorities set to 0.01 for important and 0.1
    for unimportant auxiliary variables.  The values are drawn
    for the constraints $t_{K+1} \le T$ by solid lines
    and for $t_{K+1} = T$ by dashes.}

  \label{FSM5}
\end{figure}

Figure \ref{FSM5} displays the counterpart of Figure \ref{FSM4}
for priorities in $\bp$ set, inappropriately, to 0.01 for
important and 0.1 for unimportant auxiliary variables,
to explore the sensitivity to an unobserved (and uncalibrated)
additional auxiliary variable.  The plots in the two panels have
similar features to their counterparts in Figure \ref{FSM4},
but with two important differences.
The values of $\Delta\htheta$ and $\Delta\htheta^{(0)}$
are much greater
in this uncongenial setting of inappropriate values of $\bp$
and the anomaly of decreasing values $\Delta\htheta$ with the constraint
$t_{K+1} = T$ occurs in the same range of values of $T$.
The ratios $\Delta\htheta / \Delta\htheta^{(0)}$
increase from 5.32 for $t_{K+1} = 1000$ to 30.1
for $t_{K+1} = 10\,000$.
They are higher and increase a bit more steeply than
in the congenial setting of Figure \ref{FSM4} (from 4.30 to 23.5).

\begin{figure}

  \includegraphics[width=140mm]{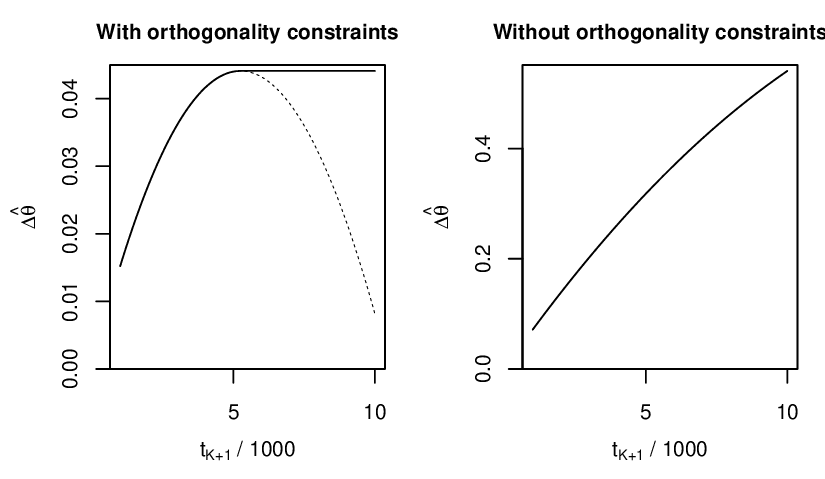}
  
  \caption{Outcomes weakly dependent on $\bX$; see text for details.
    $\Delta\htheta$ and $\Delta\htheta^{(0)}$ as functions of $t_{K+1}$
    for priorities set to 0.1 for important and 0.01
    for unimportant auxiliary variables.  The values are drawn
    by solid lines for the constraint $t_{K+1} \le T$ 
    and by dashes for $t_{K+1} = T$.}

  \label{FSM6}
\end{figure}

Next, we evaluate $\Delta\htheta$ and $\Delta\htheta^{(0)}$
for an outcome variable generated by the same process as described
earlier, but with coefficients reduced five fold, to
0.2 (instead of 1.0) for important variables.
Figure \ref{FSM6} presents the results using the established layout.
The sensitivity to a new auxiliary variable is reduced substantially,
for instance, to $\Delta\htheta = 0.044$ and
$\Delta\htheta^{(0)} = 0.318$ for $t_{K+1} = 5000$.
\begin{figure}

\includegraphics[width=140mm]{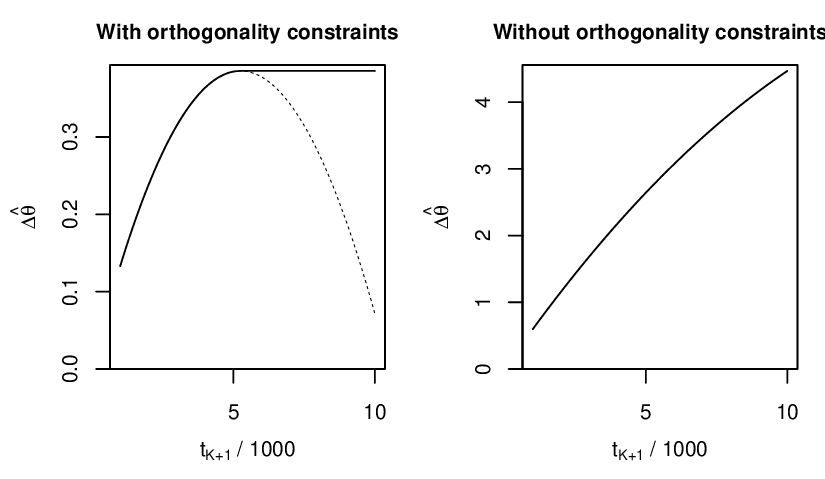}
  
  \caption{Outcomes weakly dependent on $\bX$, as in Figure \ref{FSM6},
    with inappropriate priorities $\bp$ (Figure \ref{FSM5}).}

  \label{FSM7}
\end{figure}  

If we use the same inappropriate priorities $\bp$ as in Figure \ref{FSM5}
the values of $\Delta\htheta$ and $\Delta\htheta^{(0)}$ are inflated
about nine-fold but the functions retain the same shape as
in Figure \ref{FSM6}; see Figure \ref{FSM7}.

\subsection*{Dependence on $R$}

\begin{figure}

  \includegraphics[width=140mm]{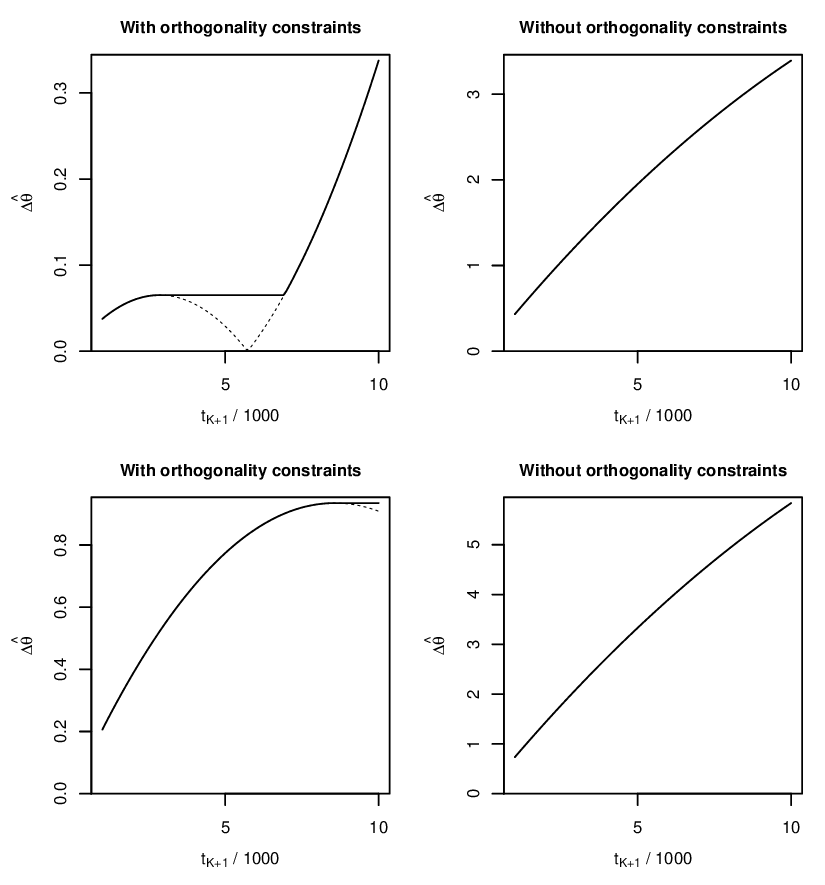}
  
  \caption{The extreme plausible changes $\Delta\htheta$
    and $\Delta\htheta^{(0)}$
    as functions of $t_{K+1}$ for $R=0.25$ (top panels)
    and  $R=0.75$ (top panels).}

  \label{FSM8}
\end{figure}

Figure \ref{FSM8} illustrates the dependence of $\Delta\htheta$
and $\Delta\htheta^{(0)}$
on $t_{K+1}$ for $R=0.25$ (top panels) and $R=0.75$ (bottom)
for the original setting introduced in Figure \ref{FSM4}.
With increasing $R$ the anomaly for the constraint $\bX\tra\bx_{K+1} = \bzer$
moves to the right
Without the orthogonality constraints, the values
of $\Delta\htheta^{(0)}$ also increase
with $t_{K+1}$ and have similar shapes. 

\begin{figure}

  \includegraphics[width=140mm]{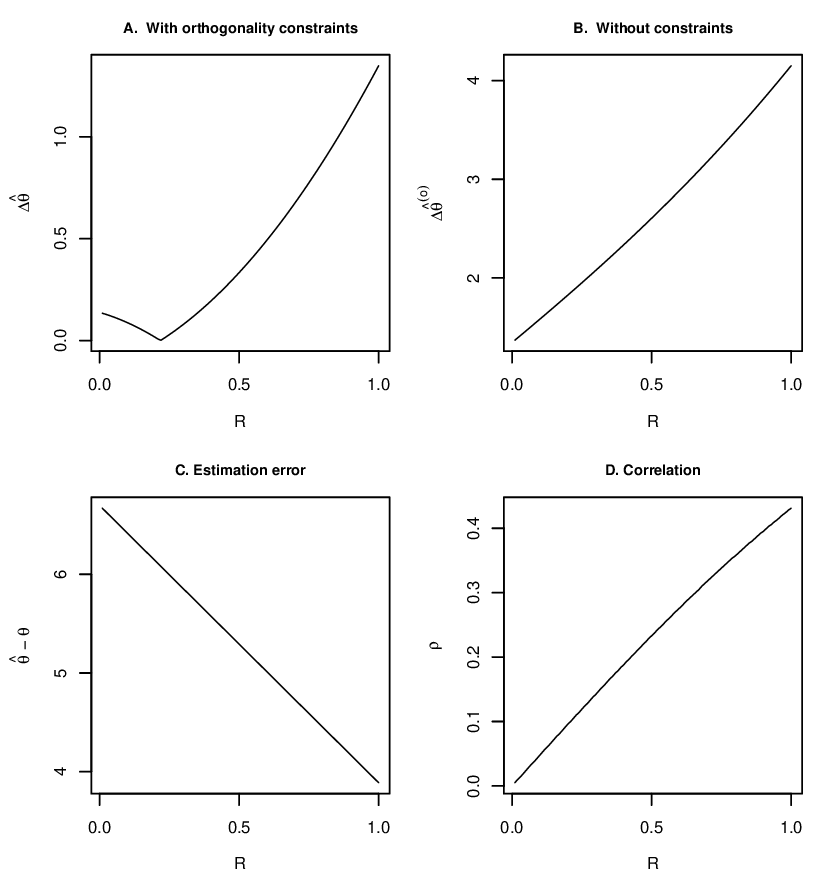}
  
  \caption{Dependence of $\Delta\htheta$, $\Delta\htheta^{(0)}$,
    estimation error and correlation of $\bx_{K+1}$ and $\bx_{K+1}^{(0)}$
    on the priority parameter $R$.}

  \label{FSMA}
\end{figure}

Figure \ref{FSMA} summarises the dependence of the sensitivity of $\htheta$
on $R$.  Panel A displays the largest plausible change
$\Delta\htheta$ as a function of $R \in (0, 1)$ with the standard setting
of $\bp$.  The largest change is quite small for $R=0$, equal to 0.134;
it decreases for $R \in (0, 0.21)$ to 0.007 and then increases, reaching
1.35 at $R=1$.  Without the orthogonality constraints (panel B),
the largest plausible change is uniformly larger,
increasing from 1.38 at $R=0$ to 4.15 at $R=1$;
$\Delta\htheta(R)$ is close to the linear function $1.38 + 2.77R$.
However, the error of the calibration estimate, $\htheta - \theta$,
decreases with $R$, approximately linearly as $6.67 - 2.68R$.
The correlation of the extreme plausible variables $\bx_{K+1}$
and $\bx_{K+1}^{(0)}$ is very small at $R=0$ and increases,
without much curvature, throughout $R$, approximately as $0.05 + 0.426R$.

\subsection*{Dependence on $\bp$}

\begin{figure}

  \includegraphics[width=140mm]{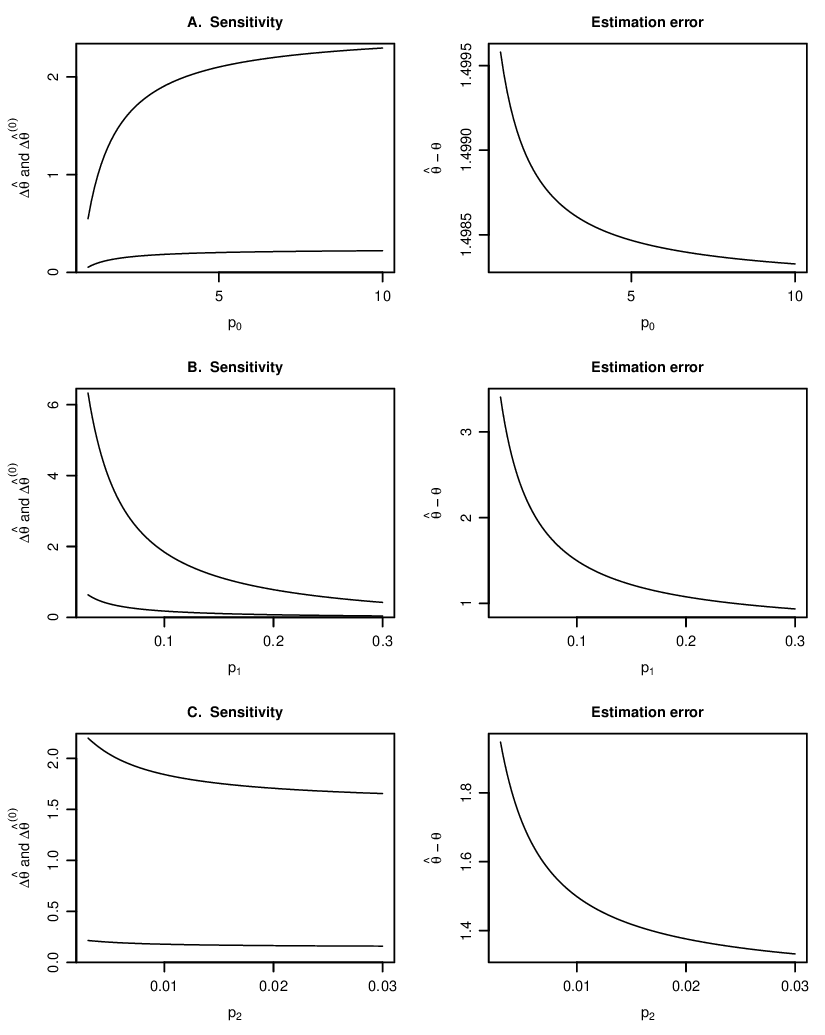}
  
  \caption{Dependence of the extreme plausible values
    of $\Delta\htheta$ and $\Delta\htheta^{(0)}$
    and estimation error on $p_{0\,}$, the common priority
    for the important auxiliary variables ($p_1$), and the common
    priority for the unimportant variables ($p_2$).  In each left-hand plot,
    the curve for $\Delta\htheta$ is above the curve
    for $\Delta\htheta^{(0)}$.}
  \label{FSMC}
\end{figure}

Figure \ref{FSMC} shows how sensitivity depends on the parameters
involved in $\bp$.
Row A displays $\Delta\htheta$ and $\Delta\htheta^{(0)}$
as functions of $p_0$ in the range $(1, 10)$.
Sensitivity of both estimates increases with $p_{0\,}$, more steeply
for small values of $p_{0\,}$.  The values of all the other priority
parameters are held at their `standard' values, e.g., $R=0.5$. 
The right-hand panel displays the plot of the error
of the calibration estimate, $\htheta - \theta$, as a function of $p_{0\,}$.
The estimate is affected by $p_0$ only slightly.
The panels in row B display the sensitivity with respect to the priority
of the important variables, that is, $p_{1\,}, p_{3\,}, \ldots, p_{11\,}$,
for $p_1 \in (0.03, 0.30)$. 
The values of $\Delta\htheta$ and $\Delta\htheta^{(0)}$
decrease with $p_{1\,}$ more steeply
for small values of $p_{1\,}$.
The panels in row C display the sensitivity with respect to the priority
of the unimportant variables ($p_{2\,}, p_{4\,}, \ldots, p_{12}$)
for for $p_2 \in (0.003, 0.030)$.

In all three cases, the functions $\Delta\htheta$
and $\Delta\htheta^{(0)}$,
as well as the estimation errors $\htheta - \theta$ appear to have asymptotes.
So, sensitivity is a concern mainly for small values of priorities. 

\subsection*{D. Simulations. Recovery of an omitted
auxiliary variable.}

In this section we use a simpler setting to study how well an omitted
auxiliary variable is recovered by re-calibration.
We generate a population of $N = 5000$ units with auxiliary variables
$\bX_{1\,}$, $\bX_2$ and $\bX_{3\,}$, each of them a random sample
from the standard normal distribution ${\cal{N}}(0,1)$,
and an outcome variable as $\bY = \bX_1 + \bX_2 + \bX_3 + \bepsi$,
where $\bepsi$ is a random sample from ${\cal{N}}(0, \sigma^2)$;
the four random samples are mutually independent.

We apply the Poisson sampling design with all the probabilities
set to 0.04, so that the expected sample size is 200.
The sample versions of $\bX_{1\,}$, $\bX_{2\,}$, $\bX_3$ and $\bY$
are denoted by the corresponding lowercases.
We set first $\sigma^2$ to 0.04 and 
calibrate the sample on $\bx_1 $ and $\bx_2$ and assess
how well $\bx_3$ is recovered by re-calibration.
We measure the ability to recover $\bx_3$ by its correlation
with the extreme plausible variable, in the main text also denoted
by $\bx_{3\,}$.  To distinguish the two variables,
we refer to the re-calibration variable by $\bx_{3\,}^{\ast}$.
The correlation with $\bx_3$ is recorded also for $\bx_3^{(0)}$,
the extreme plausible new variable obtained by re-calibration
without the orthogonality constraints.
The simulations use 1000 replications.

The importance parameters are set to $\bp = (5, 0.1, 0.1)$ and $R = 0.5$,
and the sensitivity parameters to $p_3 = 0.1$ and $t_3 = 210$.
In experiments we alter these values one at a time
and vary the residual variance of $(y_{\,}|_{\,}\bX)$ from zero to 0.85.
With $\sigma^2 = 0.04$, the omitted auxiliary variable is recovered
very well in every replication.
The sample correlations of $\bx_3^{\ast}$ with $\bx_3$
have mean (standard deviation) 0.977 (0.019) and the correlations
of $\bx_3^{(0)}$ with $\bx_3$ are 0.975 (0.008).  The two re-calibration
variables, $\bx_3^{\ast}$ and $\bx_3^{(0)}$,
are extremely highly correlated, 0.998 (0.001), confirming 
that the principal source of uncertainty is the `new' auxiliary information,
which is recovered by re-calibration very well.

Next we calibrate on all three variables, setting
$\bp = (5, 0.1, 0.1, 0.1)$ and retain the sensitivity parameters
$p_4 = 0.1$ and $t_4 = 210$.  
The mean correlation of $\bx_4^{(0)}$ with $\bx_3$ is 0.228 (0.019)
and the mean correlation of $\bx_4^{\ast}$ with $\bx_4^{(0)}$
is 0.917 (0.014).   So, the uncertainty shifts a bit toward $\bp$
but a `new' auxiliary variable is found nevertheless,
even though the model by which $\by$ was generated
suggests that there should be none. 

\begin{figure}[t]

  \includegraphics[width=140mm]{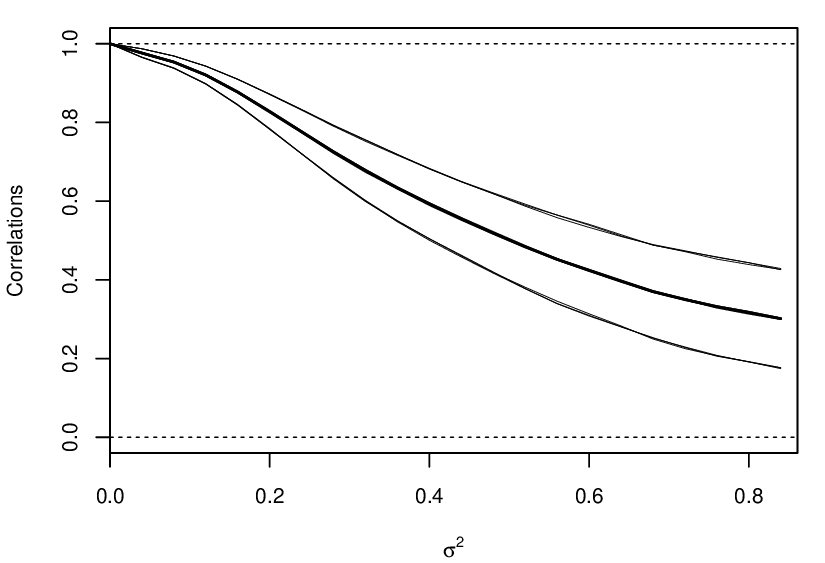}
  
  \caption{Mean correlation of the extreme plausible new auxiliary
    variable $\bx_3^{\ast}$ with the omitted variable $\bx_{3\,}$;
    re-calibration with the orthogonality constraints. Setting
    with $\sigma^2$ and $t_3 = 21$, 210 and 2100.  The hairlines
    mark the symmetric 95\% range of the replicate values.}
  \label{Fcor1}
\end{figure}

We repeat this simulation of calibrating on $\bx_1$ and $\bx_2$
for a grid of values of $\sigma^2$ in $(0, 0.84)$ and
values of $t_3$ set to 21, 210 and 2100. 
Figure \ref{Fcor1} displays the plot of the mean correlations, $M$
(solid lines), and $M \pm 2S$, hairlines,
where $S$ is the standard deviation of the replicate correlations.
The plot shows that the extreme plausible variable $\bx_3^{\ast}$
depends on $t_3$ only slightly (or not at all) and the mean correlation
is a decreasing function of $\sigma^2$.
For greater $\sigma^2$, $\bx_3^{\ast}$ is less closely related to $\bx_{3\,}$.
The studied correlations do not depend on $p_{3\,}$. 

We have also established by simulations that the correlations
of $\bx_3^{\ast}$ with $\bx_3$ depend on $p_3 \in (0.01, 0.25)$
very weakly.  For all settings, the results with and without
the orthogonality constraints differ imperceptibly --
the uncertainty about the new auxiliary information
dominates the uncertainty about the importance parameters. 
\vspace*{5mm}

\subsection*{E. Software}

All the computing, for the examples and simulation in Section 4,
as well as to check the formulae in Section 3, was conducted
in {\tt R}.  The core code for generating a population,
drawing a sampling from it, calibrating the sample and
conducting the sensitivity analysis is reproduced here.

{
  \baselineskip 15pt \small \parskip -12pt
\begin{verbatim}

##  Filename  Empir.an1

##   Generating a population
C25popR <- C25popF(popN = 120000, samN = 1000,
   bet = c(1, rep(c(1, 0.1), 6)), sig = 0.4, lmt = c(0, 25),
   betP = c(rep(0, 3), 0.35, rep(0, 5), 0.7, 0, 0.4, 0), sigP = 0)

##   The priorities
C25Pri <- list(p=c(3, rep(c(0.1, 0.01), 6)), R=0.5)

##   The sensitivity parameters
C25sns <- c(pk1=0.1, tk1=5000)

##   Drawing a sample
C25samR <- C25samF(pop = C25popR)

##   Calibration 
C25clbR <- C25clbF(samL = C25samR, Prio = C25Pri, dgt = 2)

##   Sensitivity analysis with the orthogonality constraints
C25snsR <- C25sncF(list(With=C25snsF(las=c(1,8), dgt=3), 

##   Sensitivity analysis without the orthogonality constraints
Without = C25sns0F(las=c(10,30), dgt=3)))

##   Estimation
C25estR <- C25estF(sam = C25samR, pop = C25popR, clb = C25clbR)
\end{verbatim}
}

{\tt C25popR} is a list comprising the matrix $\bX$, vector $\by$,
the vector of inclusion probabilities, the vector $\bt$ and
a summary of truncation (numbers of truncated outcomes at the limits
specified by the argument {\tt lmt}).
{\tt C25samR} is a list comprising the sample versions of $\bX$, $\by$
and base weights $\bw$ and the vector $\bt$.
{\tt C25clbR} is a list comprising the calibrated weights, labelled
by the order number in the population, and the matrix with five columns:
the HT estimate, the calibration estimate, the target ($t_k$) and
the errors of the HT and calibration estimates for the $K+1 = 13$
auxiliary variables.

{\tt C25snsR} is a list comprising the sample correlation
of $\bx_{K+1}$ with $\bx_{K+1}^{(0)}$ and a $2 \times 4$ table
with columns $\Delta \theta$, $\lambda_{2\,}$,
the number of iterations and a quantity describing the convergence,
for $\bx_{K+1}$ (row 1) and $\bx_{K+1\,}^{(0)}$.
Convergence is checked by the number of iterations being smaller
than the maximum (argument {\tt maxit}) and the convergence summary
exceeding {\tt tol}.

All the software, enabling reproduction of all the results,
tables and figures in the article 
will be available upon acceptance of the manuscript for publication 
\\
  {\tt www.sntl.co.uk/BackTi/Software/Calibr25}, 
together with comprehensive instructions.

\end{document}